# Synthesis and Pharmacology of Ester Modified (±)-*threo*-Methylphenidate Analogs

Howard M. Deutsch[*], Xiaocong Michael Ye[**] and Margaret M. Schweri[***]

[*]Corresponding author. Georgia Institute of Technology, School of Chemistry and Biochemistry, retired. howard.deutsch@mindspring.com
[**]Elan Pharmaceutical Co., retired
[***]Mercer University, School of Medicine, retired. schweri_mm@earthlink.net

## INTRODUCTION

The research described here is a continuation of an ongoing project whose aim is the synthesis and characterization of derivatives of (±)-*threo* -methylphenidate (TMP; Ritalin©) with potential in the treatment of cocaine (COC) abuse [1, 2, 3, 4, 5, 6, 7]. The reinforcing and stimulant effects of COC are thought to be due, at least in part, to the drug's ability to block dopamine (DA) re-uptake by binding to a stimulant binding site on the DA transporter located at the DA nerve terminal [8, 9]. TMP also binds to this site, as identified using the radioligand $^3$H-WIN 35, 428 (WIN) [1]. The premise of this work is that modification of the TMP molecule may result in one of two possible modes of pharmacotherapies for use in the treatment of COC abuse: (1) a classic cocaine antagonist which will block the binding of cocaine to the DA transporter, but have no intrinsic effect on the uptake of DA, or (2) a long-acting cocaine agonist/partial agonist which can be administered as a cocaine substitute with less reinforcing properties, analogous to the use of methadone for the treatment of heroin addiction.

Although COC and TMP are similar in their ability to block DA transport, they differ markedly in their activity at the serotonin (5-HT) transporter [5]. TMP has virtually no activity against 5-HT uptake, while COC is a potent 5-HT uptake blocker. Displacement of $^3$H-citalapram (CIT) binding is commonly used to measure activity at the 5-HT transporter. Measurement of the activity of these TMP derivatives against CIT binding was performed in order to determine whether the pharmacological profiles of these modified compounds more closely resemble TMP or COC.

Previous work from our laboratories has focused primarily on the effect of substituents added to the phenyl ring of TMP [1], removal of its phenyl ring [3], N-substitution [2, 10], alteration of the size of the heterocyclic ring [4], restricted rotation analogues [6], and/or modification of the ester function [5,7]. For the convenience of the reader, we have summarized much of our previous WIN binding data in Table A of the Appendix, which appears at the end of this paper. Please note that all of the compound numbers in Table A are preceded by an "M" prefix to distinguish them from the numbers assigned to the compounds described in this report, which do not have an alphabetical prefix. Other laboratories have explored the effect of replacement of the ester group with alkyl moieties [11].This work continues the examination of the effect of modification or replacement of the ester group on the TMP molecule with other functional groups, as well as substitution at the piperidine N and addition of 3,4 dichloro groups to the phenyl ring. The eleven compounds described here were quantified by their potency against WIN and CIT binding, as well as their effect on DA uptake.

## RESULTS AND DISCUSSION

## PHARMACOLOGY

WIN BINDING

The WIN binding data (and all other biological data) for the eleven TMP derivatives are shown in Table 1, arranged in order of decreasing potency at the WIN binding site. Potency is expressed as $IC_{50}$ (the concentration of a test compound that inhibits control radioligand binding or uptake by 50%). Data for TMP (**M39**) and COC are included for comparison purposes.

**Table 1. Structures and Biological Testing of (±)-*Threo*-methylphenidate Analogs**

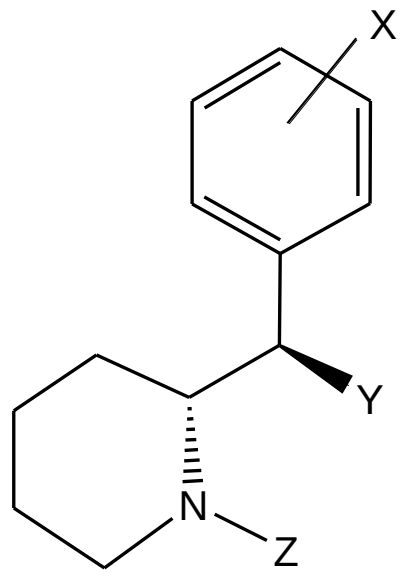

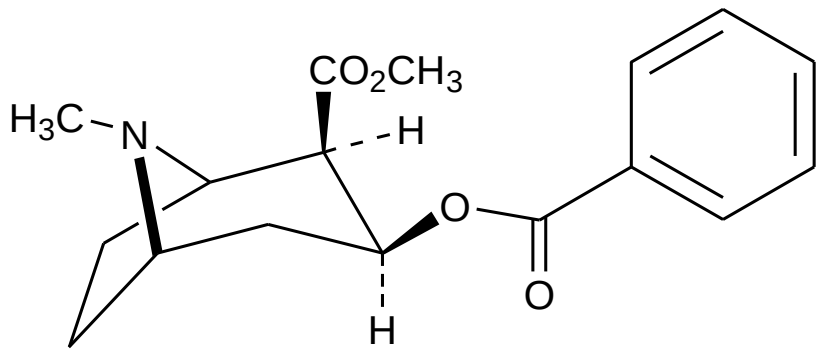

(-)Cocaine

| Cpd No. | X | Y | Z | WIN | | CIT | | DA |
|---|---|---|---|---|---|---|---|---|
| | | | | $IC_{50}$ (nM) | $n_H$ [a] | $IC_{50}$ (nM) | $n_H$ [a] | $IC_{50}$ (nM) |
| COC | | | | 160 ± 15 | 1.03 ± 0.01 | 401 ± 27 | 1.27 ± 0.01 | 404 ± 26 |
| TMP | H | $CO_2CH_3$ | H | 84.3 ± 6.3 | 0.92 ± 0.07 | >>10,000 | -- | 230 ± 16 |
| **1** | H | $CH_2OCH_2CH_3$ | benzyl | 27.0 ± 2.0 | 1.58 ± 0.01 | <<10,000 | -- | 54.5 ± 1.5 |
| **2** | 3,4-diCl | $CH_2OCH_3$ | $CH_3$ | 28.0 ± 2.0 | 1.16 ± 0.04 | 468 ± 120 | 1.29 ± 0.01 | 66.5 ± 1.5 |
| **3** | H | *t*-$CH{=}CHCO_2CH_3$ | H | 40.0 ± 2.0 | 0.94 ± 0.07 | >10,000 | - | 120 ± 6.5 |
| **4** | H | *t*-CH=CHPh | H | 40.2 ± 2.2 | 0.92 ± 0.04 | 6232 ± 86 | 1.00 ± 0.19 | 79.5 ± 7.5 |
| **5** | 3,4-diCl | $CH_2OH$ | $CH_3$ | 64.2 ± 1.2 | 1.01 ± 0.00 | 382 ± 12 | 1.27 ± 0.10 | 220 ± 21 |
| **6** | H | $CH_2OCH_2CH_3$ | H | 106 ± 1.0 | 0.98 ± 0.02 | 3,740 (1) | 0.76 (1) | 221 ± 21 |
| **7** | H | $CH{=}CH_2$ | H | 109 ± 12 | 0.95 ± 0.03 | >10,000 | -- | 242 ± 15 |
| **8** | H | $CH_2CH_2CO_2CH_3$ | H | 168 ± 11 | 0.88 ± 0.01 | >10,000 | -- | 374 ± 14 |
| **9** | H | $CH_2CH_2Ph$ | H | 193 ± 10 | 1.06 ± 0.06 | 3,160 ± 390 | 1.25 ± 0.05 | 340 ± 20 |
| **10** | 3,4-diCl | $=CH_2$ | $CH_3$ | 645 ± 42 | 1.15 ± 0.05 | 411 (1) | 1.21 | 3,350 ± 340 |
| **11** | H | $CH_2OH$ | $CH_3$ | 7,030 ± 270 | 1.21 ± 0.16 | >10,000 | -- | 12,700 ± 2,600 |

[a] $n_H$ = the Hill coefficient of binding

Three of these compounds, (**1**, **2** and **6**), have ether groups in place of the ester function of TMP. Ethyl ether **6** is approximately equipotent to TMP (see Table 2 for comparison of potencies) and is only modified at the ester position. This confirms earlier work that showed that the corresponding methyl ether of TMP (**M68**, $IC_{50}$=97.1 nM) [7] was little different than TMP. Compounds **1** and **2** have three times greater affinity for the WIN binding site than TMP itself, but are both substituted at the piperidine nitrogen, whereas TMP is a secondary amine. In addition, **2** is a 3,4-dichloro analog. Direct comparisons are difficult.

There are three compounds, **7**, **3**, and **4**, with a double bond at position Y instead of the ester group. Respectively, the double bonds are unsubstituted, or have a methyl ester or phenyl group at the second carbon. These compounds are only modified at the ester position. Ethylene analog **7** is essentially as active as TMP, confirming earlier work that nonpolar groups at this position do not lower potency [11]. Interestingly, **3** and **4**, with an ester or phenyl group, respectively, are twice as potent as TMP at the WIN binding site. When the double bonds of **3** and **4** are saturated to yield **8** and **9** there is a 4-fold loss in affinity (2-fold less potent than TMP). Thus, the nonpolar methyl ester or phenyl groups help affinity when *trans* attached to the double bond, but lower affinity in the less constrained saturated compounds. Obviously, there are some complicated steric considerations at this position.

Two compounds, **5** and **11,** have primary alcohol groups in place of the ester function of TMP. Both are N-substituted with methyl groups and in addition **5** is a 3,4-dichloro analog. Previous work showed that the conversion of the ester to alcohol [7] (**M30**) o r N-methylation [5] (**M77**) of TMP lowered affinity by 5 to 6-fold. If these effects are additive, then **11** might be expected to be 25 to 30-fold less active than TMP, whereas it is actually 83-fold less active, considerably more than additive. However, the effect of N-methyl substitution can vary as can be seen if we compare **2** to the previously reported compound (**M62**) without the N-methyl (16-fold loss) [7]. Interestingly, **5** is about as active as TMP. This shows the large effect (110-fold increase comparing **5** and **11**) on affinity that 3,4-dichloro substitution can have. This "3,4-dichloro substitution effect" has been studied in detail [6], and the results in this paper are perfectly consistent.

In contrast to N-methyl substitution, N-benzyl substitution enhanced affinity for the WIN binding site (N-benzyl ethyl ether **1** is 4-fold more potent than **6,** the ethyl ether with no substitution on the N). This is consistent with previously published data [5, 7, 10] showing that N-benzyl substitution of TMP derivatives without modification of the phenyl ring of the parent compound enhances affinity for the WIN binding site anywhere from 1.6 (for TMP itself [**M29**]) to 19-fold (for **M66,** the primary alcohol analog of TMP). See Table A, Appendix.

The most unique compound in this study is **10** which has a $sp^2$ hybridized carbon directly attached to position Y. It is formally not a *threo* compound as there is only one asymmetric carbon. Obviously the geometry in this area of the molecule would be very different than TMP. The best comparison would be to 3,4-dichloro N-methyl TMP, but this compound has not been synthesized. Closely related is 3-chloro N-methyl TMP (**M41**), which shows an $IC_{50}$ of 160 nM [5]. (3,4-dichloro TMP **M52** is equipotent to 3-chloro TMP **M50**). Using **M41** as the model compound, **10** is only 4-fold less potent than the corresponding *threo* methyl ester.

The data reported here allow further examination of another characteristic feature sometimes observed with 3,4-dichloro derivatives of the phenyl ring: a monotonic increase in the Hill coefficient ($n_H$) of WIN binding from one (denoting no positive cooperativity) to two (suggesting that there may be two interacting WIN binding sites). To aid in this discussion, $n_H$ values for the compounds to be compared are reported in Table 2. The complete selection from which the compounds of interest have been culled is shown in Table A, Appendix, along with their $n_H$ values and $IC_{50}$'s against WIN binding. This doubling of the $n_H$ was first noted with the 3,4-dichloro TMP itself [1] (**M52** in Table 2). It does not happen in all compounds with 3,4 dichloro substitutions of the phenyl ring, however, indicating that several factors may contribute to the ability of these derivatives to bind in the proper orientation to induce positive cooperativity. None of the newly synthesized dichloro compounds reported in this paper exhibit this feature (**2**, **5** and **10**), but comparison of their Hill coefficients with available data from related compounds yields some emerging patterns that may merit further investigation. Examination of the accumulated data shows that the increase of the Hill coefficient to 2 occurs upon chlorination of the TMP molecule when the ester is retained and no N-substitution is introduced, as mentioned above. It does *not* occur with dichlorinated derivatives having an unsubstituted piperidine amine, but in which the ester group is replaced by an ether, alcohol, or amide (Table 2, **M62**, **M61**, and **M72**, respectively). The Hill coefficient also remains at unity when the ester group is replaced in the dichlorinated derivative by one of the above functional groups *and* a methyl group is added to the N (Table 2: **2**, **5** and **10**). Strikingly, however, if the N-Me group is replaced by an N-benzyl group in these dihalogenated derivatives, the Hill coefficient rises to two whether the ester function is present (Table 2: **M69**) , or has been replaced by an ether (**M71**) or an alcohol (**M70**) . N-benzyl substitution in the absence of the phenyl ring chlorine substitutions is not sufficient by itself to induce positive cooperativity, as evidenced by **M29**, **M67**, and **M66**, which have Hill coefficients of one.

**Table 2. Effect of Functional Groups on Hill Coefficient of WIN Binding**

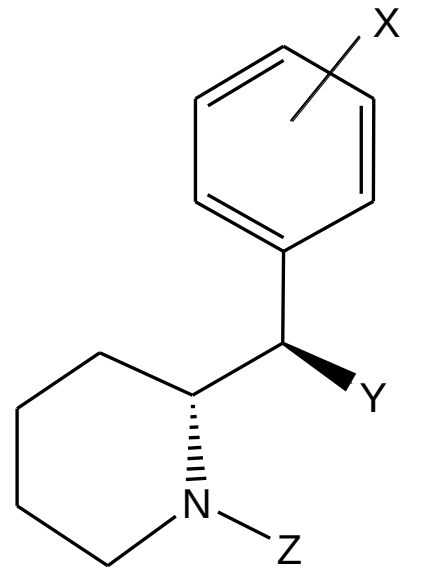

| Cpd No. | X | Y | Z | WIN $n_H$ [a] |
|---|---|---|---|---|
| TMP | H | $CO_2CH_3$ | H | 0.92 ± 0.07 |
| **M52** | 3,4-Cl | $CO_2CH_3$ | H | 2.07 ± 0.05 |
| **M29** | H | $CO_2CH_3$ | Bn | 1.08 ± 0.02 |
| **M69** | 3,4-Cl | $CO_2CH_3$ | Bn | 2.06 ± 0.20 |
| | | | | |
| **M62** | 3,4-Cl | $CH_2OCH_3$ | H | 1.30 ± 0.13 |
| **2** | 3,4-Cl | $CH_2OCH_3$ | $CH_3$ | 1.16 ± 0.04 |
| **M67** | H | $CH_2OCH_3$ | Bn | 1.12 ± 0.13 |
| **M71** | 3,4-Cl | $CH_2OCH_3$ | Bn | 2.23 ± 0.32 |
| | | | | |
| **M30** | H | OH | H | 1.07 ± 0.08 |
| **M61** | 3,4-Cl | OH | H | 1.08 ± 0.05 |
| **11** | H | OH | $CH_3$ | 1.21 ± 0.16 |
| **5** | 3,4-Cl | OH | $CH_3$ | 1.01 ± 0.00 |
| **M66** | H | OH | Bn | 1.08 ± 0.06 |
| **M70** | 3,4-Cl | OH | Bn | 1.88 ± 0.28 |
| | | | | |
| **10** | 3,4-Cl | $=CH_2$ | $CH_3$ | 1.15 ± 0.05 |
| | | | | |
| **M72** | 3,4-Cl | $CONH_2$ | H | 0.99 ± 0.11 |

[a] $n_H$ values expressed as mean ± SEM

Perusal of Table A, Appendix, shows that some compounds that are not dichlorinated also exhibit $n_H$'s of 2. We have reported previously that, in unchlorinated compounds, the chain length connecting a phenyl group to the piperidine N is critical with respect to engendering a $n_H$ of 2 [12]. Thus, in the series of N-substituted analogs (N-$[CH_2]_n$Ph; n=1-6) with either the original ester (**M29**, **M8**, **M6**, **M9**, **M10**, **M11** in order of ascending chain length) or an alcohol (**M66**, **M5**, **M2**, **M4**) at position Y, a maximum $n_H$ of ~2 is obtained when a propyl group connects the aromatic substituent to the N (**M6** and **M2**).

<u>DA UPTAKE</u>

With the exception of **10**, potency of these compounds against DA uptake generally parallels that against WIN binding, exhibiting discrimination ratio (DR; ratio of $IC_{50}$ for DA uptake to $IC_{50}$ for WIN binding) values of between 1.8 and 3.4 (Table 3). This consistent difference in affinity is most likely attributable to differences in assay conditions (*e.g.*, WIN binding is determined at $0^o$ C and equilibrium, while DA uptake is measured at $37^o$ C under non-equilibrium conditions). As with WIN binding, **1** was most potent in blocking DA uptake, while **11** was least potent. The one outlier in this group was **10**, which had a DR of 5.2. Although the reason for this is not clear, it is interesting to note that this is the only compound in this series with a planar configuration at the Y position, thus resulting in both restricted rotation and a very different conformational profile than the other *threo* compounds.

CIT BINDING

In general, these compounds had low affinity for the CIT binding site on the 5HT transporter, with $IC_{50}$'s ranging from 380 nM to >>10,000 nM. Table 3 shows the selectivity (ratio $IC_{50}$ against CIT/ $IC_{50}$ against WIN) for all compounds for which CIT binding was determined. Whereas TMP has very high selectivity for WIN (>>120), COC is fairly nonselective (*i.e.*, it is only 2.5-fold more potent at the WIN binding site than at the CIT binding site). Only one TMP analog, **10**, was truly nonselective (0.6) and was actually slightly more potent against CIT binding than WIN binding. The low affinity compound **11** likely had low selectivity (>1.4), but a precise determination could not be made. Where comparisons could be made, the remaining compounds manifested moderate to great selectivity for the WIN binding site over the CIT binding site, ranging from 6 to >250.

Some trends were evident. When the ester of TMP was replaced with a substituted double bond as in **3** and **4**, high selectivity was maintained (>250 and 155, respectively). When the double bond of these compounds was reduced yielding **8** and **9**, the selectivity was lowered (>60 and 16, respectively). The unsubstituted double bond of **7** also retained high selectivity (>92).

**Table 3. Relative Affinity, Selectivity and Discrimination Ratio (DR) for TMP Analogs**

| Compound No. | WIN | | Selectivity | DR |
|---|---|---|---|---|
| | $IC_{50}$ (nM) | Relative affinity compared to TMP | $IC_{50}$ against CIT / $IC_{50}$ against WIN | $IC_{50}$ against DA / $IC_{50}$ against WIN |
| COC | 160 | 0.53 | 2.5 | 2.5 |
| TMP | 84.3 | 1.0 | >>120 | 2.7 |
| **1** | 27.0 | 3.1 | ND | 2.0 |
| **2** | 28.0 | 3.0 | 17 | 2.4 |
| **3** | 40.0 | 2.1 | >250 | 3.0 |
| **4** | 40.2 | 2.1 | 155 | 2.0 |
| **5** | 64.2 | 1.3 | 6.0 | 3.4 |
| **6** | 106 | 0.80 | 35 | 2.1 |
| **7** | 109 | 0.77 | >92 | 2.2 |
| **8** | 168 | 0.50 | >60 | 2.2 |
| **9** | 193 | 0.44 | 16 | 1.8 |
| **10** | 645 | 0.13 | 0.6 | 5.2 |
| **11** | 7,030 | 0.012 | >1.4 | 1.8 |

While quite potent at the WIN binding site, **2** had relatively lower selectivity (17). Comparing this to **5**, which only differs from **2** in being an alcohol rather than a methyl ether, revealed considerably lower selectivity (6). Finally compared to TMP, the ethyl ether analog **6** is much less selective (>120 and 35, respectively).

The four compounds with N-methyl groups, **2**, **5**, **10**, and **11**, all had much lower selectivity than TMP (17, 6, 0.6, >1.4 and >>120, respectively).

The increase in affinity for the WIN binding site described above that was observed in the 3,4-dichloro analogs of TMP derivatives was also observed for CIT binding. In this instance, **5** was >26-fold more potent than **11** at the CIT binding site.

## CHEMISTRY

Most of the reactions used in this work were routine and the results are as expected. The one exception was the elimination of the hydroxyl group of **M61** under basic conditions to give **10**. A possible mechanism for this reaction involves attack of the alkoxide on the solvent to form the intermediate **13** as shown in Fig 1, which would have an enhanced leaving group.

**Figure 1. Possible mechanism for elimination of hydroxyl group under basic conditions**

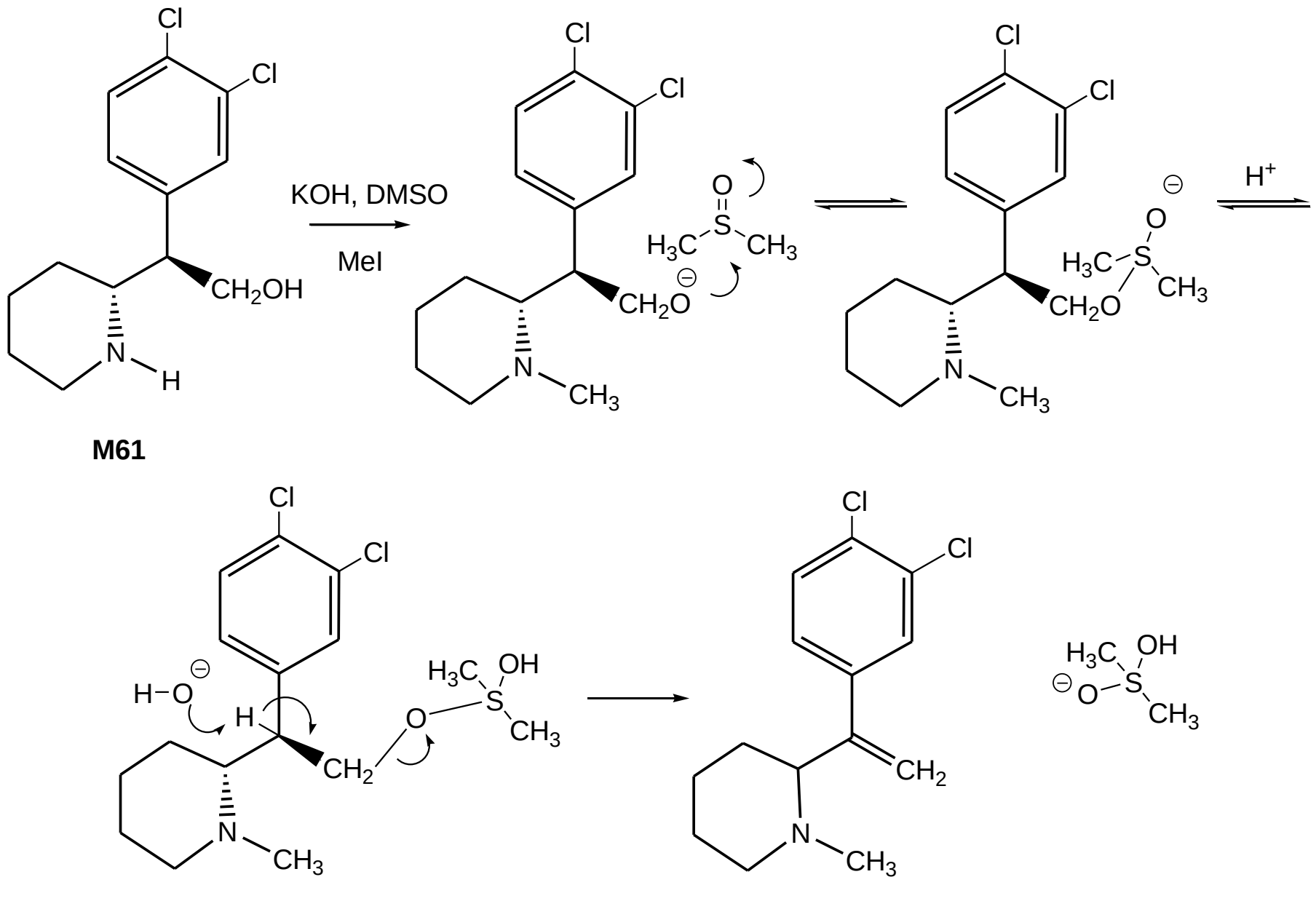

**M61**

**13** **10**

## MATERIALS AND METHODS

PHARMACOLOGICAL TESTING OF EXPERIMENTAL COMPOUNDS

WIN and CIT binding and DA uptake were determined in rat brain preparations, as described previously [1, 5] .

SYNTHESIS OF EXPERIMENTAL COMPOUNDS

All of the compounds were synthesized using standard methods of organic chemistry. All were fully characterized including: quantitative elemental, $^1$H-NMR, $^{13}$C-NMR, MS, IR analyses and in some cases X-ray. For reasons beyond the authors' control, the $^1$H-NMR, $^{13}$C-NMR, MS, IR and X-ray data were lost before this publication was complete. These data fully confirmed the purity, structure and the stereochemistry (*threo* and *trans* double bond in some cases) of all of the compounds, but cannot be presented. The syntheses of compounds **1** - **11** are summarized in Schemes 1-4 and full experimental details and quantitative elemental data are presented for all compounds in the following Experimental Procedures section.

EXPERIMENTAL PROCEDURES

General Methods. Starting materials and solvents were purchased from either Aldrich Chemical Co. or Fisher Scientific and used without further purification. Flash chromatography was run using 230-400 mesh silica gel. Melting points were determined on a Mel-Temp apparatus and are uncorrected. $^1$H and $^{13}$C spectra were obtained on a Varian Gemini-300 Spectrometer. High-resolution mass spectra [EI (electron ionization), CI (chemical ionization) or FAB (fast atom bombment)] were recorded on a VG Analytical 70-SE mass spectrometer equipped with a 11-250J data system. Elemental analyses were obtained from Atlantic Microlabs, Atlanta, GA. Free bases were dissolved in EtOAc or MeOH and converted to HCl salts by the addition of 1M HCl (1.5 equiv.) in diethylether. The excess HCl was removed under reduced pressure and the solid was recrystallized from various mixtures of MeOH and EtOAc. Nomenclature of compounds: TMP is methyl 2-phenyl-2-(piperidin-2-yl)acetate, ritalinic acid is 2-phenyl-2-(piperidin-2-yl)acetic acid and ritalinol is 2-phenyl-2-(piperidin-2-yl)ethanol. All structures are shown in Table 1 and Schemes 1-4.

Scheme 1 Experimental

**(±)-*threo*-N-(benzyl)ritalinol ethyl ether (1)**.

A mixture of 248 mg of ground KOH and 5 mL of DMSO was stirred at 55 $^oC$ for 20 min under an $N_2$ atmosphere. The alcohol **M66** (250 mg, 0.84 mmol) in 5 mL of DMSO was added followed by 110 uL (1.47 umol) of EtBr. Stirring was continued overnight, 80 mL of $CH_2Cl_2$ was added, the mixture was washed 5x20 mL of water and the organic layer dried with MgSO4. Evaporation of the solvent gave 284 mg of yellow liquid. Silica gel chromatography using 1% EtOAc in hexane gave 159 mg (58%) of pure **1**. A sample, (45.6 mg), was converted to the HCl salt, 52 mg, mp 238-240 $^oC$. Anal: Calc for $C_{22}H_{30}ClNO$; C 73.41, H 8.40, N 3.89, Cl 9.85. Found; C 73.55, H 8.60 N 3.81 Cl 9.70.

**(±)-*threo*-ritalinol ethyl ether (6)**.

A mixture of 88.5 mg of **1**, 10 mL of MeOH and 25 mg of $Pd(OH)_2$ (20%, wet) was stirred under a $H_2$ atmosphere (65 psi) for 2.5 h. Filtration and evaporation of the solvent gave 58.1 mg (94%) of pure **6**. The HCl salt was too hydroscopic to purify. Anal: Calc for free base $C_{15}H_{23}NO$; C 77.21, H 9.93, N 6.60. Found; C 76.97, H 9.98 N 6.29.

Scheme 1

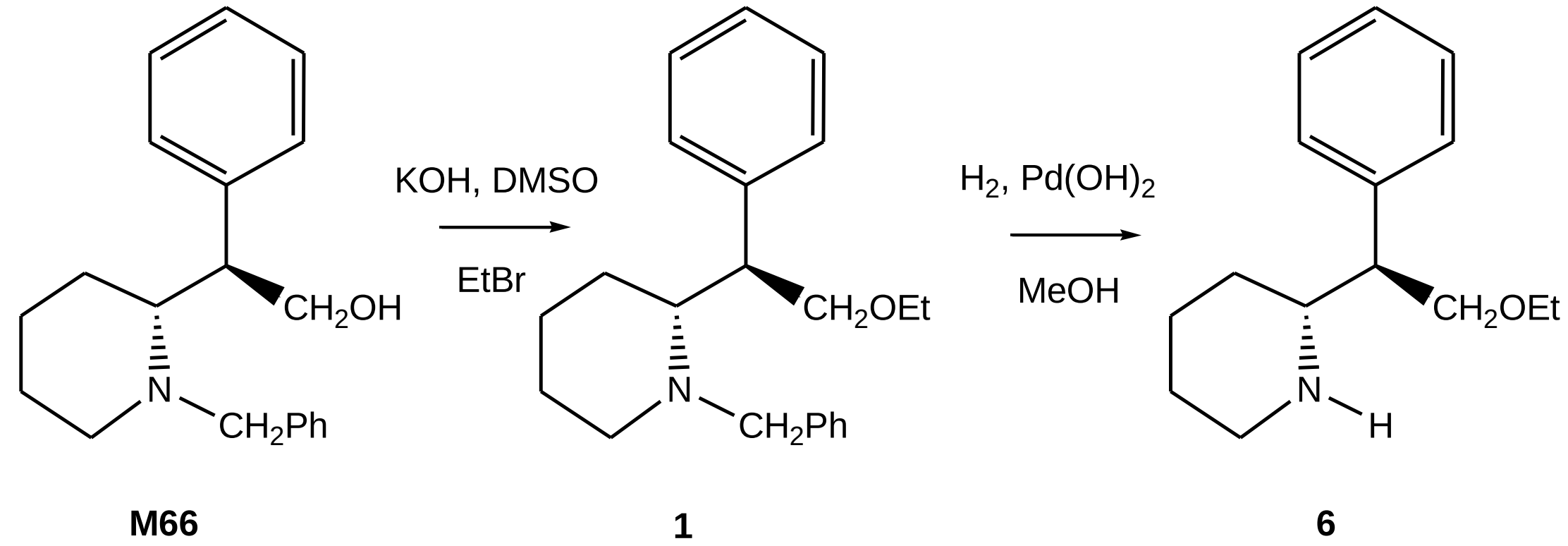

Scheme 2

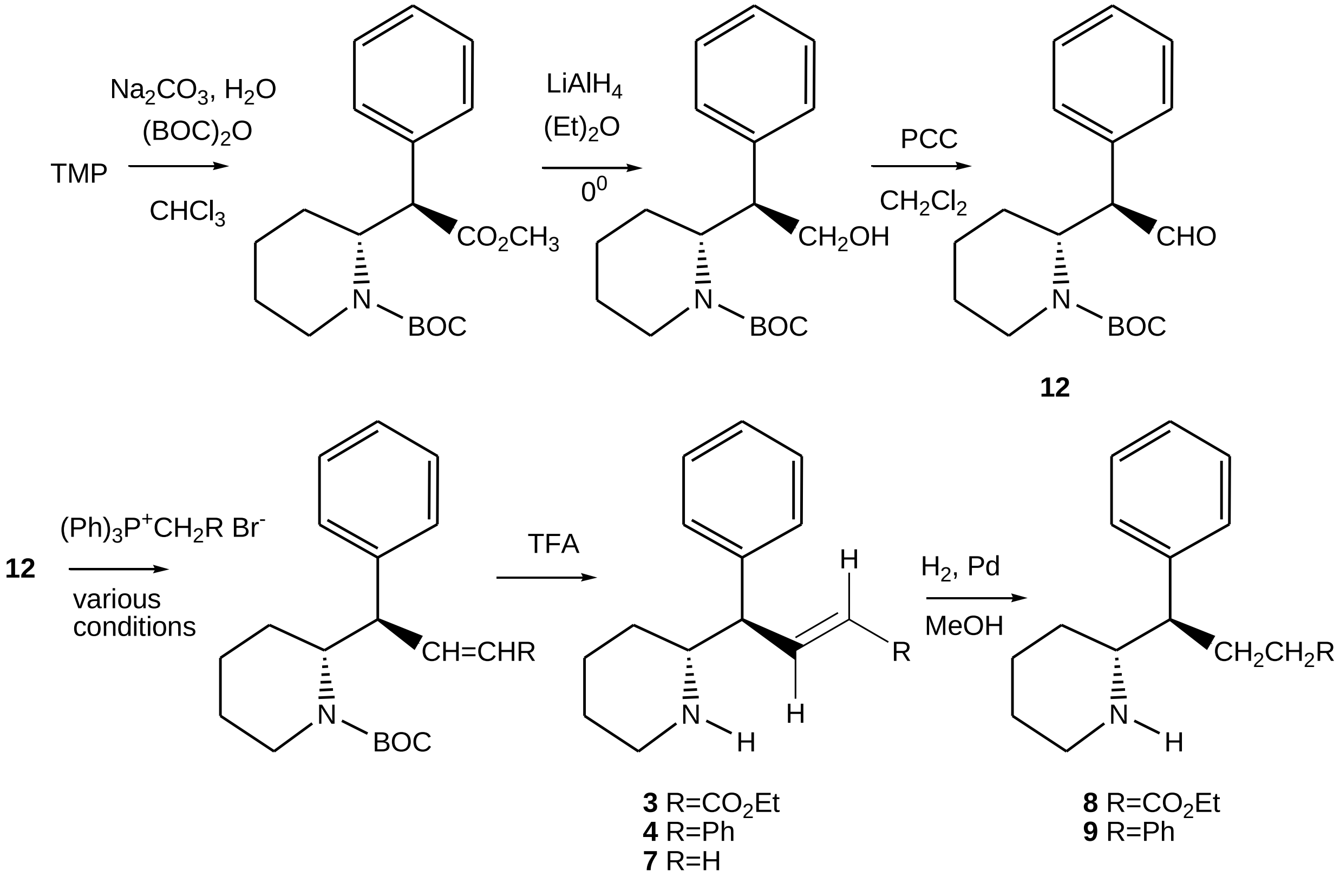

Scheme 3

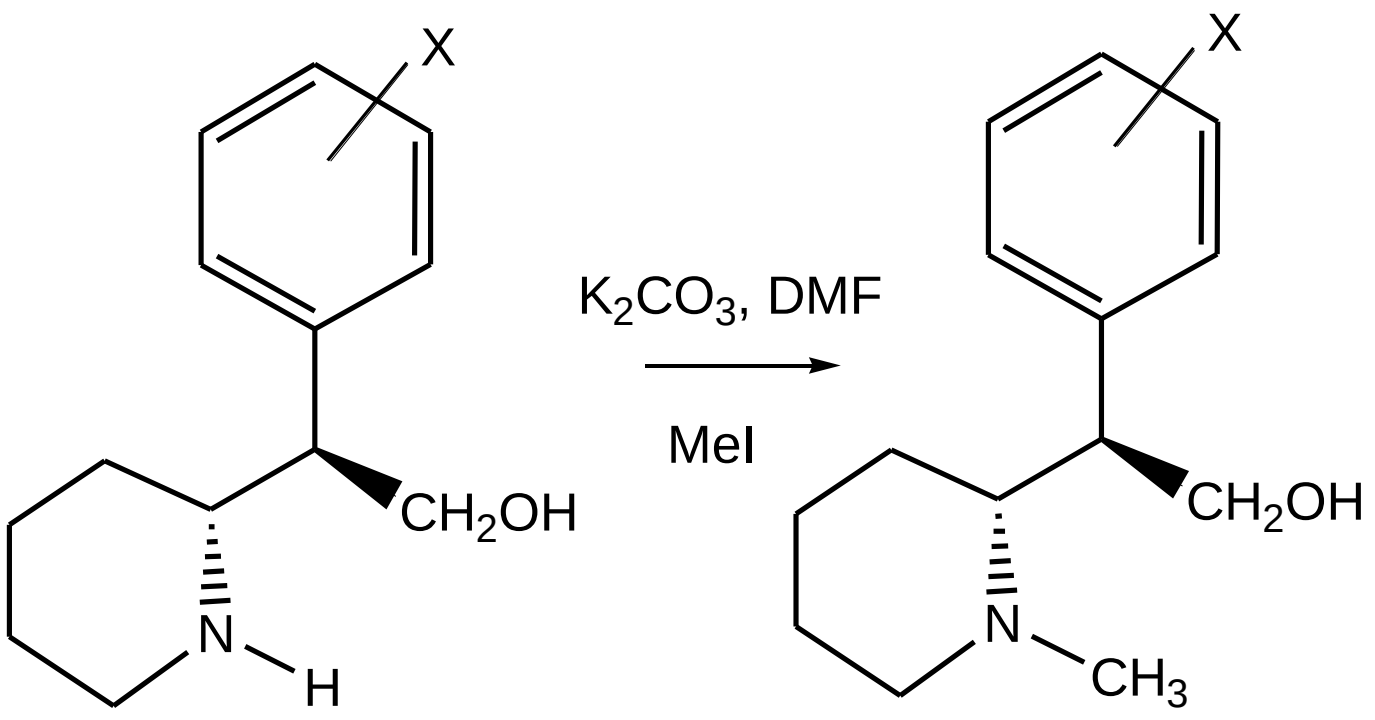

**M30** X=H
**M61** X=3,4-diCl

**11** X=H
**5** X=3,4-diCl

Scheme 4

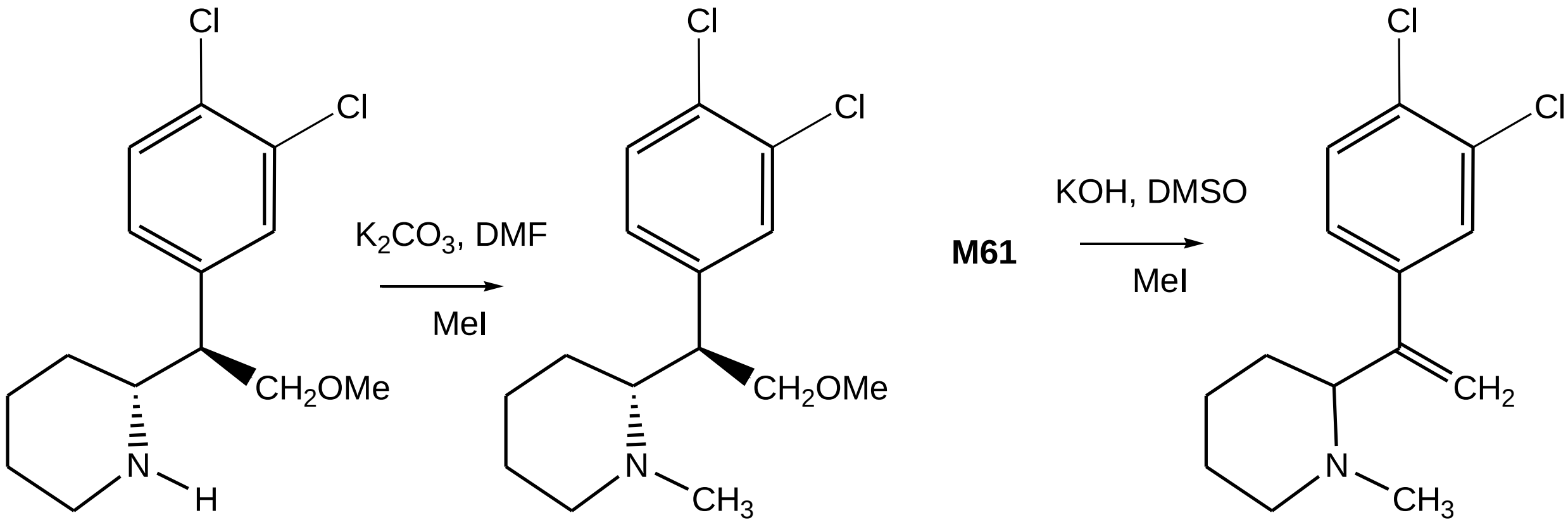

**M62**

**2**

**10**

<u>Scheme 2 Experimental</u>

**(±)-*threo*-(E)-methyl 4-phenyl-4-(piperidin-2-yl)but-2-enoate (3)**.

A mixture of 1.46 g TMP (5.40 mmol), 40 mL of $CHCl_3$, 20 mL H2O, 0.84 g $NaHCO_3$, 1.0 g NaCl and 1.39g $(BOC)_2O$ was heated under reflux for 2 h. An additional 30 mL of $CHCl_3$ was added and the organic layer washed with 1N HCl (2x25 mL) and then water. Drying and evaporation gave 2.01 g of pure product with a small excess of $(BOC)_2O$. This product was dissolved in 30 mL of $Et_2O$, cooled to 0 $^oC$ and 6.6 mL of 1.0M $LiAlH_4$ added. After stirring for 1 h, the temperature was allowed to rise to RT over 1 h, water was carefully added followed by 100 mL of $Et_2O$. The organic layer was washed with water, dried and evaporated. This gave 1.60 g (97%) of the N-BOC protected alcohol as a pure, white solid, mp 101-102 $^oC$. A portion (419 mg), dissolved in 6 mL of $CH_2Cl_2$, was added to a suspension of PCC (443 mg) in 12 mL of $CH_2Cl_2$ and stirred for 2 h at RT. Filtration and evaporation gave 365 mg of nearly pure aldehyde **12**. This material was dissolved in 12 mL of $CH_2Cl_2$ and 1.2 eq of $(Ph)_3P{=}CHCO_2CH_3$ added and then stirred at RT for 36 h. After evaporation, the oil was dissolved in 5% EtOAc/hexane and cooled to remove some of the $(Ph)_3P{=}O$. The soluble material was chromatographed on silica using 20% EtOAc/hexane to give 480 mg (65 %) of N-BOC ester. This material was mixed with 6 mL of TFA, stirred for 40 min, evaporated and treated with EtOAc and $NaCO_3$ to give after drying and evaporation, 325 mg (94%) of pure **3**. This material was converted to the HCl salt, which gave perfect crystals from EtOAc/MeOH, mp 243-244 $^{o}$C. Both $^{1}$H-NMR and single crystal Xray analysis showed the it was the *trans* (E) isomer. Anal: Calc for $C_{16}H_{22}ClNO_2$; C 64.97, H 7.50, N 4.74, Cl 11.99. Found; C 65.11, H 7.59 N 4.74 Cl 12.06.

**(±)-*threo*-(E)-2-1,3-diphenylallyl)piperidine (4).**
To a mixture of 275 mg of $(Ph)_3P^+$-$CH_2$Ph $Br^-$ and NaH (0.55 mmol) in 5 mL of $CH_2Cl_2$ was added 138 mg (0.46 mmol) of **12**. After stirring 5 days, 30 mL of EtOAc was added, washed with water, dried and evaporated to give N-BOC product. Chromatography with 2 % EtOAc in hexane yielded 134 mg (85%) of pure product. Treatment with TFA as described above, gave 98 mg (100%) of **4**. The HCl salt was crystallized from EtOAc/MeOH to give pure **4**, mp 279-281 $^{o}$C. $^{1}$H-NMR analysis showed that it was the *trans* (E) isomer. Anal: Calc for $C_{20}H_{24}ClN$; C 76.54, H 7.71, N 4.46, Cl 11.30. Found; C 76.42, H 7.71, N 4.44, Cl 11.42.

**(±)-*threo*-2-(1-phenylallyl)piperidine (7).**
NaH (132 mg) in 1.6 mL of DMSO was heated to 75 $^{o}$C for 40 min, cooled to RT and 1.18 g (2.33 mmol) of $(Ph)_3P^+$-$CH_2$ $Br^-$ in 4 mL of DMSO added. After 10 min, 200 mg of **12** was added and after stirring overnight, 20 mL of water was added and extracted with 5x40mL of hexane. Drying and evaporation gave the N-BOC product. Chromatography with 2 % EtOAc in hexane yielded 155 mg (78%). Treatment with TFA as described above, gave 88 mg (88 %) of **7**. The HCl salt was crystallized from EtOAc/MeOH to give pure **7**, mp 180-182 $^{o}$C. Anal: Calc for $C_{14}H_{20}ClN$; C 70.72, H 8.48, N 5.89, Cl 14.01. Found; C 70.86, H 8.58, N 5.84, Cl 14.83.

**(±)-*threo*-2-(1,3-diphenylpropyl)piperidine (9).**
A mixture of 100 mg of **4**, 10 mL of MeOH and 25 mg of Pd (5% on C) was stirred under a $H_2$ atmosphere (45 psi) for 2 h. Filtration and evaporation of the solvent gave 78 mg of pure **9**. The HCl salt was prepared. Anal: Calc for $C_{20}H_{26}NCl$; C 76.05, H 8.30, N 4.45, Cl 11.22. Found; C 76.40, H 8.32, N 4.31, Cl 11.04.

**(±)-*threo*-methyl 4-phenyl-4-(piperidin-2-yl)butanoate (8).**
Using the same procedure as in the preparation of **9**, 155 mg of **3** gave 141 mg of pure **8**. The HCl salt showed mp 192-193 $^{o}$C. Anal: Calc for $C_{16}H_{24}NO_2Cl$; C 64.53, H 8.12, N 4.70, Cl 11.90. Found; C 64.42, H 8.22, N 4.69, Cl 12.01.

<u>Scheme 3 Experimental</u>
**(±)-*threo*-N-(methyl)ritalinol (11)**.
A mixture of 413 mg **M30**, 570 mg of $K_2CO_3$,14 uL of MeI (1.1 eq) and 10 mL of DMF was stirred at RT for 22 h under a $N_2$ atmosphere. $(Et)_2O$ (70 mL) was added, the mixture was washed 5x25 mL of water and the organic layer dried with MgSO4. Evaporation of the solvent followed by silica gel chromatography using 1% EtOAc in hexane gave 246 mg of a pure oil. A solid HCl salt could not be prepared. Anal: Calc for $C_{14}H_{21}NO$; C 76.67, H 9.65, N 6.39. Found; C 76.39, H 9.86 N 6.19.

**(±)-*threo*-3,4-dichloro-N-(methyl)ritalinol (5)**.
Using the same procedure as in the preparation of **11**, 123 mg of **M61** gave 141 mg of solid **5**, which was recrystallized from EtOAc/hexane (4/96) and showed mp 101-102.6 $^{o}$C. Anal: Calc for $C_{14}H_{19}Cl_2NO$; C 58.34, H 6.64, N 4.86, Cl 24.60. Found; C 58.55, H 6.70 N 4.82, Cl 24.78.

Scheme 4 Experimental

**(±)-*threo*-3,4-dichloro-N-(methyl)ritalinol methyl ether (2)**.

Using the same procedure as in the preparation of **11**, 121 mg of **M62** gave 83 mg of pure **2**, as an oil. The HCl salt was too hydroscopic to isolate. Anal: Calc for $C_{15}H_{21}Cl_2NO$; C 59.61, H 7.00, N 4.86, Cl 23.46. Found; C 59.83, H 7.09, N 4.51, Cl 23.10.

**2-(1-(3,4-dichlorophenyl)vinyl)-1-methylpiperidine (10)**

A mixture of 125 mg of ground KOH and 15 mL of DMSO was stirred at RT for 20 min under a $N_2$ atmosphere. The alcohol **M61** (126 mg, 0.46 mmol) in 5 mL of DMSO was added followed by 87 uL (3 eq) of MeI. Stirring was continued overnight, 60 mL of $CH_2Cl_2$ was added, the mixture was washed 5x20 mL of water and the organic layer dried with MgSO4. Evaporation of the solvent gave 125 mg of yellow liquid, which was obviously a mixture by TLC. Silica gel chromatography using 1% EtOAc in hexane gave 28 mg of pure **10**. Anal: Calc for $C_{14}H_{17}Cl_2N$;C 62.23, H 6.34, N 5.18, Cl 26.24. Found; C 61.82, H 6.73, N 4.79 Cl 25.88.

## CONCLUSIONS

All compounds in this study are modified at the ester function on the TMP molecule. Other modifications include dichlorination of the phenyl ring and/or substitution of the piperidinyl nitrogen. These changes have significant effects on its activity against WIN and CIT binding, as well as synaptosomal DA transport.

The synthesis and testing of the compounds reported here suggest several avenues for further exploration. Compounds **1** through **4** are of interest because they are significantly more potent than TMP itself. Of these four compounds, **3** merits special attention because of its very high selectivity for the WIN over the CIT binding site. Thus, its pharmacological profile should more closely resemble that of TMP rather than COC. In comparison, **5** is slightly more active than TMP, but much less selective, and might be somewhat more COC like.

Even though less potent than TMP, **10** has several features that recommend it for further study. Its rigid planar conformation at position Y (what was an asymmetric carbon of the parent compound) could make it a useful tool for elucidating the structural details of the stimulant binding site. This compound also exhibits the highest DR of this series. If this finding reflects an actual difference in the activity of this compound against DA uptake relative to WIN binding, it may prove effective in blocking COC binding at doses that have a lesser effect on DA uptake, making it useful as a treatment for COC addiction. Moreover, it is the only compound of this series with no selectivity for the WIN over the CIT binding site: the modifications incorporated into the molecule caused it to lose potency for the WIN binding site while it gained potency at the CIT binding site, compared to TMP. This characteristic could make it a useful tool for the comparison of the neurotransmitter uptake sites on the DA and 5HT transporters.

The three dichlorinated compounds reported here (**2**, **5**, and **10**), combined with halogenated and other TMP derivatives previously described by us [1, 5, 7, 12] should yield useful clues as to the positive cooperativity that can be induced at the WIN binding site. This is a relatively unexplored area, but one that holds great promise for advancing our understanding of the DA transporter.

# APPENDIX

## Table A. WIN Binding for Selected Methylphenidate Analogs

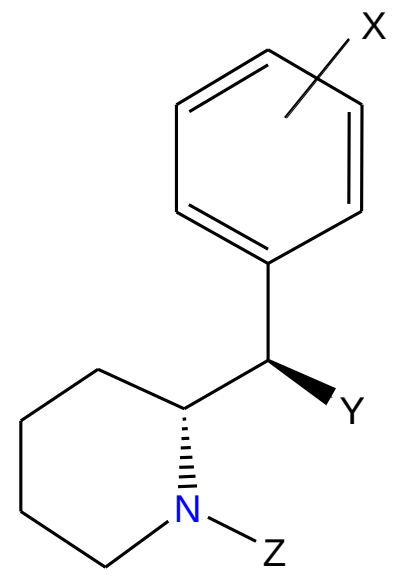

| Compound [a] | X [b] | Z [b] | Y [b] | $IC_{50}$ (nM) [c] | $n_H$ [c,d] |
|---|---|---|---|---|---|
| **M1** | -H | 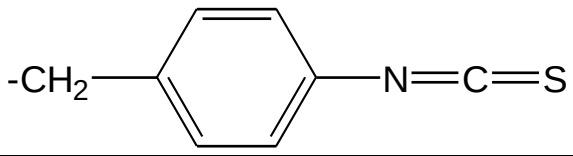 | $-CO_2CH_3$ | 422 ± 13 | 1.46 ± 0.06 |
| **M2** | -H | $-(CH_2)_3Ph$ | $-CH_2OH$ | 194 ± 14 | 1.92 ± 0.01 |
| **M3** | -H | $-CH_2C \equiv CH$ | $-CO_2CH_3$ | 821 ± 100 | 1.10 ± 0.20 |
| **M4** | -H | $-(CH_2)_4Ph$ | $-CH_2OH$ | 623 ± 64 | 1.54 ± 0.08 |
| **M5** | -H | $-(CH_2)_2Ph$ | $-CH_2OH$ | 1430 ± 150 | 1.09 ± 0.02 |
| **M6** | -H | $-(CH_2)_3Ph$ | $-CO_2CH_3$ | 267 ± 13 | 1.78 ± 0.05 |
| **M7** | -H | 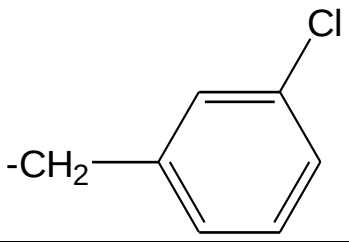 | $-CO_2CH_3$ | 106 ± 24 | 1.36 ± 0.16 |
| **M8** | -H | $-(CH_2)_2Ph$ | $-CO_2CH_3$ | 678 ± 46 | 1.21 ± 0.02 |
| **M9** | -H | $-(CH_2)_4Ph$ | $-CO_2CH_3$ | 205 ± 44 | 1.34 ± 0.18 |
| **M10** | -H | $-(CH2)_5Ph$ | $-CO_2CH_3$ | 1570 ± 80 | 1.27 ± 0.07 |
| **M11** | -H | $-(CH_2)_6Ph$ | $-CO_2CH_3$ | 656 ± 17 | 0.95 ± 0.05 |
| **M12** | -H | 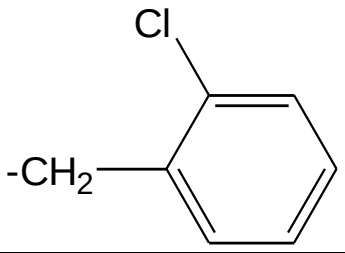 | $-CO_2CH_3$ | 243 ± 40 | 1.19 ± 0.09 |
| **M13** | -H | $-CH_2CH=CH_2$ | $-CO_2CH_3$ | 597 ± 4 | 0.97 ± 0.04 |
| **M14** | -H | 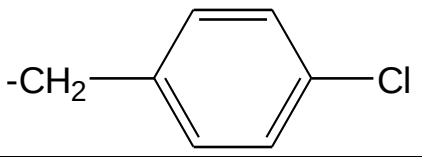 | $-CO_2CH_3$ | 31.2 ± 5.7 | 1.45 ± 0.12 |
| **M15** | -H | 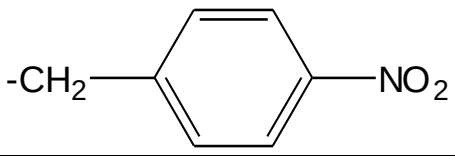 | $-CO_2CH_3$ | 113 ± 3 | 1.02 ± 0.07 |
| **M16** | -H | 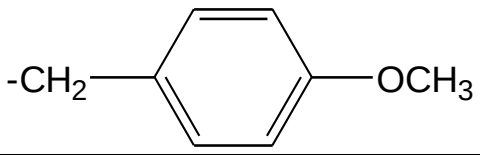 | $-CO_2CH_3$ | 79.1 ± 1.4 | 1.21 ± 0.03 |
| **M17** | -H | 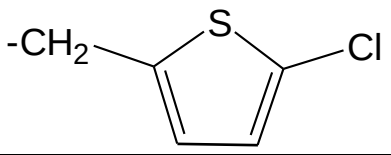 | $-CO_2CH_3$ | 392 ± 15 | 2.31 ± 0.26 |
| **M18** | -H | 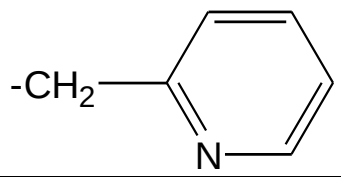 | $-CO_2CH_3$ | 369 ± 4 | 1.10 ± 0.08 |
| **M19** | -H | 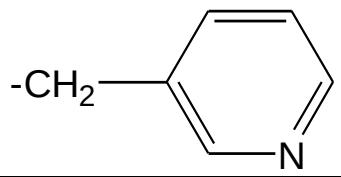 | $-CO_2CH_3$ | 173 ± 15 | 1.07 ± 0.04 |
| **M20** | -H | 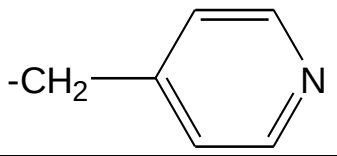 | $-CO_2CH_3$ | 128 ± 13 | 1.16 ± 0.01 |
| **M21** | -H | 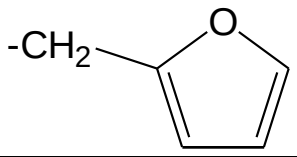 | $-CO_2CH_3$ | 536 ± 38 | 1.06 ± 0.10 |
| **M22** | -H | 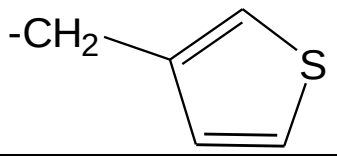 | $-CO_2CH_3$ | 143 ± 25 | 1.01 ± 0.00 |
| **M23** | -H | 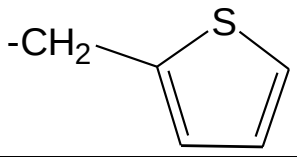 | $-CO_2CH_3$ | 224 ± 1 | 1.25 ± 0.05 |
| **M24** | -H | 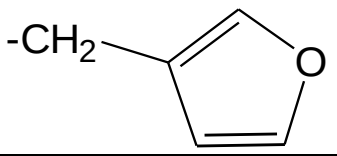 | $-CO_2CH_3$ | 459 ± 67 | 1.08 ± 0.01 |
| **M25** | -H | $-CH_2CH_3$ | $-CH_2OH$ | 2340 ± 780 | 0.85 ± 0.06 |
| **M26** | 3’,5’-di$CH_3$ | -H | $-CO_2CH_3$ | 4690 ± 60 | 0.96 ± 0.05 |
| **M27** | 3’,5’-diCl | -H | $-CO_2CH_3$ | 65.6 ± 5.4 | 1.16 ± 0.01 |
| **M28** | -H | 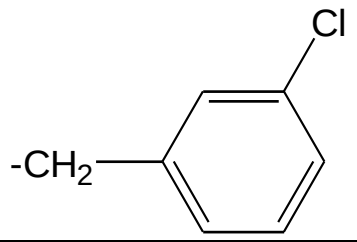 | $-CH_2OH$ | 25.8 ± 0.2 | 1.21 ± 0.04 |
| **M29** [f] | -H | $-CH_2Ph$ | $-CO_2CH_3$ | 52.9 ± 2.3 | 1.08 ± 0.02 |
| **M30** | -H | -H | $-CH_2OH$ | 448 ± 8 | 1.07 ± 0.08 |
| **M31** [g] | 4’-OH | -H | $-CO_2CH_3$ | 98.0 ± 10 | 1.07 ± 0.12 |
| **M32** | 3’-$CH_2OH$,<br>4’-$OCH_2OH$ | $-CH_3$ | $-CO_2CH_3$ | 620 ± 40 | 1.04 ± 0.02 |
| **M33** | 4’-OH | $-CH_3$ | $-CO_2CH_3$ | 1220 ± 140 | 1.05 ± 0.01 |
| **M34** [g] | 4’-$NO_2$ | -H | $-CO_2CH_3$ | 494 ± 33 | 1.18 ± 0.10 |
| **M35** [g] | 3’-$NH_2$ | -H | $-CO_2CH_3$ | 265 ± 5 | 1.06 ± 0.13 |
| **M36** [g] | 4’-$NH_2$ | -H | $-CO_2CH_3$ | 34.5 ± 4.0 | 0.96 ± 0.09 |
| **M37** [g] | 4’-$OCH_3$ | -H | $-CO_2CH_3$ | 83.0 ± 11 | 0.83 ± 0.10 |
| **M38** [g] | 4’-Cl | -H | $-CO_2CH_3$ | 20.6 ± 3.4 | 1.17 ± 0.09 |
| **M39** [g] | -H | -H | $-CO_2CH_3$ | 83.0 ± 7.9 | 0.90 ± 0.09 |
| **M40** [g] | 4’-t-butyl | -H | $-CO_2CH_3$ | 13500 ± 450 | 1.12 ± 0.08 |
| **M41** [h] | 3’-Cl | $-CH_3$ | $-CO_2CH_3$ | 160 ± 18 | 0.96 ± 0.04 |
| **M42** [e] | 4’-I | -H | $-CO_2CH_3$ | 14.0 ± 0.0 | 1.10 ± 0.04 |
| **M43** [e] | 3’-Br | -H | $-CO_2CH_3$ | 4.18 ± 0.17 | 1.14 ± 0.07 |
| **M44** [f] | 4’-$CH_3$ | $-CH_3$ | $-CO_2CH_3$ | 140 ± 9 | 1.02 ± 0.03 |
| **M45** [e] | 2’-Br | -H | $-CO_2CH_3$ | 1870 ± 135 | 0.93 ± 0.00 |
| **M46** [g] | 2’-$OCH_3$ | -H | $-CO_2CH_3$ | 101000 ± 10000 | 0.94 ± 0.09 |
| **M47** [g] | 3’-$OCH_3$ | -H | $-CO_2CH_3$ | 288 ± 52 | 1.11 ± 0.16 |
| **M48** [g] | 2’-OH | -H | $-CO_2CH_3$ | 23100 ± 50 | 1.04 ± 0.04 |
| **M49** [g] | 3’-OH | -H | $-CO_2CH_3$ | 321 ± 1 | 1.09 ± 0.02 |
| **M50** [g] | 3’-Cl | -H | $-CO_2CH_3$ | 5.1 ± 1.6 | 0.95 ± 0.02 |
| **M51** [g] | 4’-F | -H | $-CO_2CH_3$ | 35.0 ± 3.0 | 0.97 ± 0.02 |
| **M52** [g] | 3’,4’-diCl | -H | $-CO_2CH_3$ | 5.30 ± 0.70 | 2.07 ± 0.05 |
| **M53** [g] | 3’,4’-di$OCH_3$ | -H | $-CO_2CH_3$ | 810 ± 10 | 1.12 ± 0.00 |
| **M54** [e] | 4’-Br | -H | $-CO_2CH_3$ | 6.90 ± 0.10 | 1.15 ± 0.07 |
| **M55** [g] | 2’-Cl | -H | $-CO_2CH_3$ | 1950 ± 230 | 0.98 ± 0.02 |
| **M56** [g] | 2’-F | -H | $-CO_2CH_3$ | 1420 ± 120 | 0.90 ± 0.02 |
| **M57** [g] | 3’-F | -H | $-CO_2CH_3$ | 40.5 ± 4.5 | 0.85 ± 0.10 |
| **M58** [g] | 4’-$CH_3$ | -H | $-CO_2CH_3$ | 33.0 ± 1.2 | 1.05 ± 0.02 |
| **M59** [g] | 3'-$CH_3$ | -H | -$CO_2CH_3$ | 21.4 ± 1.1 | 1.01 ± 0.12 |
| **M60** | 3'-F | -H | -$CH_2OH$ | 281 ± 32 | 1.08 ± 0.05 |
| **M61** | 3',4'-diCl | -H | -$CH_2OH$ | 4.20 ± 0.52 | 1.08 ± 0.05 |
| **M62** | 3',4'-diCl | -H | -$CH_2OCH_3$ | 1.70 ± 0.24 | 1.30 ± 0.13 |
| **M63** | 3'-Cl | -$CH_2Ph$ | -$CO_2CH_3$ | 41.2 ± 3.4 | 1.93 ± 0.20 |
| **M64** | -H | -$CH_2Ph$ | -$CON(CH_3)_2$ | 1730 ± 52 | 1.01 ± 0.03 |
| **M65** | -H | -$CH_2Ph$ | -$CONH_2$ | 384 ± 8 | 1.00 ± 0.02 |
| **M66** [h] | -H | -$CH_2Ph$ | -$CH_2OH$ | 23.7 ± 3.3 | 1.08 ± 0.06 |
| **M67** [h] | -H | -$CH_2Ph$ | -$CH_2OCH_3$ | 17.8 ± 1.1 | 1.12 ± 0.13 |
| **M68** | -H | -H | -$CH_2OCH_3$ | 97.1 ± 10.3 | 0.97 ± 0.04 |
| **M69** | 3',4'-diCl | -$CH_2Ph$ | -$CO_2CH_3$ | 76.3 ± 2.7 | 2.06 ± 0.20 |
| **M70** | 3',4'-diCl | -$CH_2Ph$ | -$CH_2OH$ | 2.74 ± 0.35 | 1.88 ± 0.28 |
| **M71** | 3',4'-diCl | -$CH_2Ph$ | -$CH_2OCH_3$ | 4.2 ± 1.2 | 2.23 ± 0.32 |
| **M72** | 3',4'-diCl | -H | -$CONH_2$ | 16.4 ± 2.4 | 0.99 ± 0.11 |
| **M73** | -H | -H | -$CO_2CH_2Ph$ | 1020 ± 130 | 1.04 ± 0.11 |
| **M74** | 3'-$CH_3$ | -$CH_3$ | -$CO_2CH_3$ | 108 ± 16 | 1.00 ± 0.04 |
| **M75** | -H | -H | -$CH_2O(CO)CH_3$ | 690 ± 270 | 0.91 ± 0.04 |
| **M76** | -H | -H | -$CONH_2$ | 1730 ± 170 | 0.91 ± 0.04 |
| **M77** [h] | -H | -$CH_3$ | -$CO_2CH_3$ | 499 ± 25 | 1.00 ± 0.01 |
| **M78** | 4'-$C_2H_5$ | -H | -$CO_2CH_3$ | 737 ± 78 | 0.96 ± 0.03 |
| **M79** [i] | 3',4'-benzo | -H | -$CO_2CH_3$ | 11.0 ± 2.5 | 1.01 ± 0.03 |
| **M80** [i] | 4'-$CF_3$ | -H | -$CO_2CH_3$ | 615 ± 15 | 0.97 ± 0.05 |

[a] Unless otherwise referenced in the first column, the syntheses of the compounds and WIN binding data shown here were reported previously in Misra *et al*. [7]. **M39** is the parent compound, (±)-*threo*-methylphenidate (TMP).
[b] Naming convention for X substituents is based on the six positions of the phenyl ring: e.g. **M48** (2'-OH TMP) has the -OH substituent at the 2'-position of the phenyl ring. "Benzo" refers to a 1,2-disubstituted benzene ring (3',4'-benzo=alpha naphthyl analogue). For Y and Z substituents, Ph = phenyl.
[c] Binding and $n_H$ data were generated in the laboratory of MMS. Values are expressed as the mean ± standard error of the mean (SEM) of 2-7 assays for each compound.
[d] Hill coefficient of [$^3$H]WIN 35,428 binding. Unless otherwise referenced in Column 1, these values represent previously unpublished data.
[e] Compounds **M42**, **M43**, **M45**, and **M54** were provided by Dr. S. J. Gatley of Brookhaven National Laboratories.
[f] Synthesis and binding studies have been described elsewhere. [10]
[g] Synthesis and binding studies have been described elsewhere. [1]
[h] Synthesis and binding studies have been described elsewhere. [5]
[i] Synthesis and binding studies have been described elsewhere. [4]

**ACKNOWLEDGMENTS**
This work was supported in part by NIH grants DA06305 and DA11541. MMS gratefully acknowledges the technical assistance of Ms. Monica Shields, Ms. LaKeisha Nelson, and Mr. Eric Shields.